\documentstyle[12pt]{article}
\newcommand\be{\begin{equation}}
\newcommand\ee{\end{equation}}
\newcommand\bea{\begin{eqnarray}}
\newcommand\eea{\end{eqnarray}}
\newcommand\ket[1]{|#1\rangle}

\newcommand{\fatalpha}{{\bf \alpha \kern -0.44em \alpha}}
\newcommand{\fatsigma}{{\bf \sigma \kern -0.54em \sigma}}
\newcommand{\tpchi}{{\bf \chi \kern -0.35em \chi}}
\newcommand{\llambda}{{\bf \lambda \kern -0.45em \lambda}}

\bibliography{plain}
\title{\bf Generation of quantum entanglement between three level atoms via $n$ coupled cavities}\vspace{20mm}
\author{ R. Sufiani$^{a,b}$
  \thanks{E-mail:sofiani@tabrizu.ac.ir}   and
  A. Darkhosh$^{c}$
 \thanks{E-mail:a.darkhosh@atauni.edu.tr}    ,
 \\ $^a$ {\small Department of Theoretical Physics and Astrophysics,
University of Tabriz, Tabriz 51664, Iran.} \\ $^b$ {\small Institute
for Studies in Theoretical Physics and Mathematics, Tehran
19395-1795, Iran.} \\ $^c$ {\small Physics Department, Faculty of
Science, Atat\"{u}rk University, Erzurum, Turkey.}} \pagebreak

\vspace{20mm}
\begin{document}
\maketitle \vspace{15mm}
\newpage
\begin{abstract}
Based on two-photon exchange interaction between $n$ coupled optical
cavities each of them containing a single three level atom, the
$n$-qubit and $n$-photonic state transfer is investigated. In fact,
following the approach of Ref.\cite{Alex1}, we consider $n$ coupled
cavities instead of two cavities and generalize the discussions
about quantum state transfer, photon transition between cavities and
entanglement generations between $n$ atoms. More clearly, by
employing the consistency of number of photons (the symmetry of
Hamiltonian), the hamiltonian of the system is reduced from $3^n$
dimensional space into $2n$ dimensional one. Moreover, by
introducing suitable basis for the atom-cavity state space based on
Fourier transform, the reduced Hamiltonian is block-diagonalized,
with $2$ dimensional blocks. Then, the initial state of the system
is evolved under the corresponding Hamiltonian and the suitable
times $T$ at which the initially unentangled atoms, become maximally
entangled, are determined in terms of the hopping strength $\xi$
between cavities.  \\


{\bf PACs Index: 03.65.Ud }
\end{abstract}

\vspace{70mm}
\newpage
\section{Introduction}
The quantum communication between several parts of a physical unit,
is a crucial ingredient for many quantum information processing
protocols \cite{1}. Schemes for the transfer of quantum information
and the generation and distribution of entanglement have been
designed and implemented, in the past years, in a number of physical
systems (see for example \cite{11}-\cite{20}). Atoms and ions are
particularly considered as tools for storing quantum information in
their internal states. Naturally, photons represent the best qubit
carrier for fast and reliable communication over long distances
\cite{ref5,ref6}. Recently, using photons in order to achieve
efficient quantum transmission between spatially distant atoms has
considered in several works \cite{Alex1, ref1, ref, Alex0, Alex2,
ozgur}. The basic idea, is to utilize strong coupling between
optical cavities and the atoms. On the other hand, due to the
ability of quantum entanglement as a resource for several quantum
information processing tasks such as quantum communication, and
certain quantum cryptographic protocols, the creation of quantum
entanglement naturally arises as goals in nowadays quantum control
experiments in studying the nonclassical phenomena in quantum
physics.

One of the known models in quantum optics describing the atom-field
interaction is the Jaynes-Cummings Hamiltonian \cite{book1, book2}.
In the study of three-level atoms, M. Alexanian and S. Bose
\cite{Alex0} introduced a unitary transformation, whereby the
three-level atom was reduced to a corresponding two-level atom of
the Jaynes-Cummings type with two-photon instead of single-photon
transitions. In Refs. \cite{Alex1, Alex2}, entanglement properties
of two and three atom-cavity systems in which the cavities are
coupled via two-photon exchange interaction, was analyzed in detail.
Such results could set the pathway towards massively correlated
multiphoton nonlinear quantum optical systems \cite{ref7,ref8},
which are rapidly developing modern subjects nowadays. The
motivation of interest to such systems is their promise in quantum
switching, quantum communication and computation and quantum phase
transition applications.

In this paper, following the approach of Refs.
\cite{Alex1,Alex0,Alex2}, and introducing some suitable basis in
which the Hamiltonian of the system can be block-diagonalized, we
generalize the discussions of Ref.\cite{Alex1} and Ref.\cite{Alex2}
for two and three atom-cavity systems, to a system consisting of $n$
coupled atom-cavity subsystems. More clearly, we consider a system
of $n$ spatially separated optical cavities, each containing a
single three level atom, which are coupled to each other with
two-photon exchange interaction. Our objective is to examine state
transfer (atomic state exchange or photon transition) within photon
and atom subsystems and to consider possible generation of the
particle entanglement between the subsystems.

The organization of the paper is as follows. In section 2, the model
describing a system of $n$ identical atom-cavity subsystems is
introduced. The main results of the paper such as
block-diagonalization of the Hamiltonian of the system, solving the
Shr\"{o}dinger equation for time dependent probability amplitudes of
the state of the system, and discussions about state transfer
(atomic excitation or photon transitions) and entanglement
generation between atoms or photons, are given in this section.
Sections $3$ and $4$ are respectively concerned with the special
cases of two and three identical coupled cavities. Paper is ended
with a brief conclusion.
\section{The Model: $n$ coupled cavities via two-photon exchange interaction}
We will consider $n$ identical cavities each containing one
three-level atom, where the cavities are coupled via two photon
hopping between them. In fact, we consider that the cavities are
located at the nodes of the complete graph $K_n$ with $n$ nodes and
each cavity interacts with all of the other cavities via two-photon
exchange.

Let us first introduce the two-photon Hamiltonian obtained via an
exact unitary transformation introduced in Ref. \cite{Alex0}:
$$ H^{(i)}=\hbar\omega N^{(i)}+ E^{(i)}_0+\hbar\mu \sigma^{(i)}_{ee} +\hbar\eta \sigma^{(i)}_{gg}+\hbar\lambda(\sigma^{(i)}_{eg} a^2_i +\sigma^{(i)}_{ge} a_i^{2{\dag}})$$
where, the operator
$$\hat{N}^{(i)}=a_i^{\dag}a_i+\sigma^{(i)}_{ee}-\sigma^{(i)}_{gg}+1$$
is a constant of motion for the $i$-th atom-cavity subsystem, i.e,
we have $[H^{(i)}, \hat{N}^{(i)}]=0$ for each $i=1,2,\ldots, n$. The
operators $a_i$ and $a_i^{\dag}$ are photon operators of the $i$-th
cavity, and $\sigma^{(i)}_{ab}=|a\rangle^{{{(i)}{(i)}}}\langle b|$,
for $ i = 1, 2,\ldots, n$ denote the atomic transition operators for
the $i$-th cavity referring to either the ground (g) or excited (e)
state. Now, the Hamiltonian for the $n$ cavities is given by:
\begin{equation}\label{H}
H=\sum_{i=1}^n( H^{(i)}-H_0^{(i)})+\hbar\xi \sum_{i,j=1; i<
j}^n(a_i^{2{\dag}}a_j^2+a_j^{2{\dag}}a_i^2),\end{equation} where,
$$H_0^{(i)}=\hbar\omega (\hat{N}^{(i)}-1)+(E_g+E_e)/2,$$
with $E_g$, $E_e$ being the energies of the ground and exited
states, respectively. The last term in the Hamiltonian (\ref{H}) is
the two-photon exchange interaction between the cavities,
characterized by the hopping rate $\xi$. The parameters $ E_0, \mu,
\eta$ and $\lambda$ are the free energies of the subsystems written
in the notation of Ref. \cite{Alex0} and we do not need their clear
definitions in the present paper. All of these parameters depend on
the photon number in the corresponding cavities and so, on the
cavity-mode intensity through the eigenvalues of the operator
$\hat{N}^{(i)}$.

The operator $\hat{N}=\sum_{i=1}^n {\hat{N}}^{(i)}$ commutes with
the Hamiltonian (\ref{H}) and so we can reduce the Hamiltonian to
the subspace spanned with the eigenstates of $\hat{N}$ and consider
the time evolution of the states in this subspace. For a given
eigenspace of $\hat{N}$ with eigenvalue $N$, the maximum possible
number of photons in a cavity is $N$ when the corresponding atom is
in the ground state, which occurs when there are no photons present
in the other cavities and the atoms are also in the ground state.
Then, the total number of photons in the system will be $N$. The
constant number of total photons determines the subspace or the
manifold in which the states evolve in time (the initial state of
the system determines the constant number $N$). We will consider the
manifold with $N=2$. In this case, each single atom-cavity system
can take one of the three possible states $\ket{g,0}$, $\ket{g,2}$
or $\ket{e,0}$, and so, the total possible states that the system of
$n$-cavities can take, are $3^n$ states. Due to the consistency of
total $N=2$, the only possible states which we can have, are $2n$
states instead of $3^n$ ones. In fact, these $2n$ states are
eigenstates of $\hat{N}$ with eigenvalue $2$, and the
$3^n$-dimensional Hamiltonian $H$ is reduced to $2n$-dimensional one
in the bases which span the eigenspace of $\hat{N}$ with the
corresponding eigenvalue $2$. The bases states that span this
subspace or manifold, are given by:
$$\ket{c_i}=\ket{g,0}\ldots \ket{g,0}\underbrace{\ket{g,2}}_{i-th}\ket{g,0}\ldots \ket{g,0},$$
$$\ket{a_i}=\ket{g,0}\ldots \ket{g,0}\underbrace{\ket{e,0}}_{i-th}\ket{g,0}\ldots \ket{g,0},$$
for $i=0,1,\ldots, n-1$. Indeed, these bases span the eigenspace of
$\hat{N}$ with eigenvalue $2$, i.e., we have $\hat{N} (\alpha
\ket{c_i}+\beta \ket{a_i})=2(\alpha \ket{c_i}+\beta \ket{a_i})$.
Therefore, the general time dependent state of the $n$-cavity system
is given by
\begin{equation}\label{sai}
\ket{\psi(t)}=\sum_{i=0}^{n-1}(C_i(t)\ket{c_i}+A_i(t)\ket{a_i}).
\end{equation}
Then, one can easily show that, by considering the order of bases as
$\ket{c_0}, \ket{a_0},\ldots, \ket{c_{n-1}}, \ket{a_{n-1}}$, the
Hamiltonian $H$ takes the following direct product form
\begin{equation}\label{H1}H=I_n \otimes \left(\begin{array}{cc}
               1 & \tan\theta_0 \\
                  \tan\theta_0 & \tan^2\theta_0 \\
                \end{array}\right)+2\xi (J_n-I_n)\otimes \left(\begin{array}{cc}
               1 & 0 \\
                 0 & 0 \\
                \end{array}\right),
                \end{equation}
                where, $I_n$ is $n\times n$ identity matrix and $J_n$ is the all one matrix of order $n$. The quantity $\tan \theta_0$ is given by
                $\tan \theta_0=\frac{1}{\sqrt{2}r}$ with
                $r=\frac{g_1}{g_2}$, where $g_1$ and $g_2$ are the atom-photon coupling constants in
the three-level atom. In writing the above equation, the
dimensionless time $[(E_0^+-E_0^-)\cos^2 \theta_0]t/\hbar\rightarrow
t$ and dimensionless hopping constant $\hbar\xi/[(E_0^+-E_0^-)\cos^2
\theta_0]\rightarrow\xi$ have introduced (see Ref.\cite{Alex1,
Alex2} for the cases $n=2$ and $n=3$ coupled cavities), where
$E_0^+$ and $E_0^-$ are eigenvalues associated with eigenvectors
$\ket{\psi_0^{+}}^{(i)}=\sin \theta_0\ket{e,0}+ \cos \theta_0
\ket{g,2}$ and $\ket{\psi_0^{-}}^{(i)}=\cos \theta_0\ket{e,0}+ \sin
\theta_0 \ket{g,2}$ of $H^{(i)}$, respectively.

It is well known that the matrix $J_n$ has eigenvalues $0$, and $n$
(due to the fact that $J_n^2=nJ_n$), and is diagonalized by discrete
Fourier transform $F$ defined as $F_{kl}:=
\frac{1}{\sqrt{n}}\omega^{kl}$ for $k,l=0,1,\ldots, n-1$, where
$\omega=\exp{(\frac{2\pi i}{n})}$ is the $n$-th root of unity.
Therefore, by introducing the new Fourier transformed bases
$\{\ket{c_i}', \ket{a_i}'\}_{l=0}^{n-1}$ as:
$$
\ket{c_l}':=
\frac{1}{\sqrt{n}}\sum_{i=0}^{n-1}\omega^{li}\ket{c_i},$$
\begin{equation}\label{basis}
\ket{a_l}':= \frac{1}{\sqrt{n}}\sum_{i=0}^{n-1}\omega^{li}\ket{a_i}
\end{equation}
and considering the ordering $\{\ket{c_0}', \ket{a_0}';
\ldots;\ket{c_{n-1}}', \ket{a_{n-1}}'\}$, the Hamiltonian (\ref{H1})
takes the following block diagonalized form:
\begin{equation}\label{H2}H=I_n \otimes \left(\begin{array}{cc}
               1 & \tan\theta_0 \\
                  \tan\theta_0 & \tan^2\theta_0 \\
                \end{array}\right)+2\xi\;\ diag{(n-1,-1,\ldots,-1)}\otimes \left(\begin{array}{cc}
               1 & 0 \\
                 0 & 0 \\
                \end{array}\right),
                \end{equation}
where, $diag {(n-1,-1,\ldots,-1)}$ is the $n\times n$ diagonal
matrix with diagonal entries as $n-1$ and $-1$ respectively.  Now,
by using the Schr\"{o}dinger equation of motion $i\hbar
\frac{\partial}{\partial t}\ket{\psi}=H\ket{\psi}$, the equations of
motion are given by:
$$i\dot{C'_0}=[1+2\xi(n-1)]C'_0+\tan \theta_0A'_0;$$
$$\hspace{-1.5cm}i\dot{A'_0}=\tan \theta_0C'_0+\tan^2 \theta_0A'_0,$$
and
$$
i\dot{C'_l}=(1-2\xi)C'_l+\tan \theta_0A'_l;$$
\begin{equation}\label{eq.mo}
\hspace{-0.5cm}i\dot{A'_l}=\tan \theta_0C'_l+\tan^2 \theta_0A'_l,
\end{equation}
for $ l=1,2,\ldots, n-1$. The equations (\ref{eq.mo}) can be exactly
solved for any value of $\tan \theta_0$. Here we take the ratio of
atomic couplings in the three level atoms as $r=1/\sqrt{2}$ so that
we have $\tan \theta_0=1$. Substituting $\tan \theta_0=1$ in
(\ref{eq.mo}) and solving the corresponding differential equations,
one can obtain
$$\hspace{-1cm}C'_0(t)=\frac{e^{-i[1+\xi(n-1)]t}}{\sqrt{1+\xi^2(n-1)^2}}\{[\sqrt{1+\xi^2(n-1)^2}\cos t\sqrt{1+\xi^2(n-1)^2}-i\xi (n-1)\sin t\sqrt{1+\xi^2(n-1)^2}]C'_0(0)-$$
$$i\sin t\sqrt{1+\xi^2(n-1)^2}A'_0(0)\},$$
$$\hspace{-1cm}A'_0(t)=\frac{e^{-i[1+\xi(n-1)]t}}{\sqrt{1+\xi^2(n-1)^2}}\{[\sqrt{1+\xi^2(n-1)^2}\cos t\sqrt{1+\xi^2(n-1)^2}+i\xi (n-1)\sin t\sqrt{1+\xi^2(n-1)^2}]A'_0(0)-$$
\begin{equation}\label{eq1}
i\sin t\sqrt{1+\xi^2(n-1)^2}C'_0(0)\}
\end{equation}
where, for $l=1,2,\ldots, n-1$ we obtain
$$\hspace{-1cm}C'_l(t)=\frac{e^{-i(1-\xi)t}}{\sqrt{1+\xi^2}}\{[\sqrt{1+\xi^2}\cos t\sqrt{1+\xi^2}+i\xi\sin t\sqrt{1+\xi^2}]C'_l(0)-i\sin t\sqrt{1+\xi^2}A'_l(0)\},$$
\begin{equation}\label{eq2}
\hspace{-1cm}A'_l(t)=\frac{e^{-i(1-\xi)t}}{\sqrt{1+\xi^2}}\{[\sqrt{1+\xi^2}\cos
t\sqrt{1+\xi^2}-i\xi\sin t\sqrt{1+\xi^2}]A'_l(0)-i\sin
t\sqrt{1+\xi^2}C'_l(0)\}.
\end{equation}
By using (\ref{basis}), one can obtain the time dependence of the
coefficients $C_i(t)$ and $A_i(t)$ of the state of the system in
(\ref{sai}) via the inverse Fourier transform as,
$$C_i(t)=\frac{1}{\sqrt{n}}\sum_{l=0}^{n-1}\omega^{-li}C'_l(t),$$
\begin{equation}\label{eq3}
A_i(t)=\frac{1}{\sqrt{n}}\sum_{l=0}^{n-1}\omega^{-li}A'_l(t).
\end{equation}

It should be pointed out that, one can evaluate the probabilities
associated with the state of the system as a superposition of atomic
states $\ket{a_i}$, and that of photonic states $\ket{c_i}$, denoted
by $P_a(t)$ and $P_c(t)$, respectively. For instance, considering
the initial state
$\ket{\psi(0)}=\frac{1}{\sqrt{n}}(\ket{g,2}\ket{g,0}...
\ket{g,0}+\ket{g,0}\ket{g,2}\ket{g,0}...\ket{g,0}+\ldots+
\ket{g,0}...\ket{g,0}\ket{g,2})$, with initial conditions $A_l(0)=0$
and $C_l(0)=\frac{1}{\sqrt{n}}$ for all $l=0,1,..., n-1$, with the
aid of Eqs. (\ref{sai}) and (\ref{eq2}), we obtain
$$P_c(t)=\sum_{l=0}^{n-1}|C_l(t)|^2=\sum_{l=0}^{n-1}|C'_l(t)|^2=\frac{1}{n[1+\xi^2(n-1)^2]}\{\xi^2(n-1)^2+\cos^2 t\sqrt{1+\xi^2(n-1)^2}\}+$$
$$\frac{n-1}{n(1+\xi^2)}\{\xi^2+\cos^2 t\sqrt{1+\xi^2}\},$$
\begin{equation}\label{pr}
\hspace{-8cm}P_a(t)=\sum_{l=0}^{n-1}|A_l(t)|^2=\sum_{l=0}^{n-1}|A'_l(t)|^2=1-P_c(t).
\end{equation}
where, in the second equality in $P_c(t)$ and that of $P_a(t)$, we
have used the fact that the Fourier transform is unitary and so dose
not change the norm of vectors. The above result indicates that, in
the limit of large $\xi\rightarrow \infty$, we have $P_c(t)\simeq 1$
for every time $t$, i.e., for large enough $\xi$, all of the atoms
will be at their ground state $\ket{g}$ at every time $t$.
\subsection{Large and small hopping strengths}
One should notice that for large values of the hopping strength,
i.e., $\xi\gg$, the evaluated coefficients $C'_i(t)$ and $A'_i(t)$
in (\ref{eq1}) and (\ref{eq2}) take the form
$$\hspace{0.35cm}C'_0(t)\simeq e^{-i[1+\xi(n-1)]t}\{e^{-i\xi(n-1)t}C'_0(0)-\frac{i\sin \xi(n-1)t}{\xi(n-1)}A'_0(0)\},$$
$$A'_0(t)\simeq e^{-i[1+\xi(n-1)]t}\{e^{i\xi(n-1)t}A'_0(0)-\frac{i\sin
\xi(n-1)t}{\xi(n-1)}C'_0(0)\},$$
$$\hspace{-3cm}C'_l(t)\simeq e^{-i(1-\xi)t}\{e^{i\xi t}C'_l(0)-\frac{i\sin \xi t}{\xi}A'_l(0)\},$$
\begin{equation}\label{eq1'}
\hspace{2cm}A'_l(t)\simeq e^{-i(1-\xi)t}\{e^{-i\xi
t}A'_l(0)-\frac{i\sin \xi t}{\xi}C'_l(0)\};\;\;\;\ l=1,2,\ldots,
n-1.
\end{equation}
Neglecting also the second terms in the above approximations, we get
$$\hspace{-2cm}C'_0(t)\approx e^{-2i\xi(n-1)t}C'_0(0),$$
$$C'_l(t)\approx e^{2i\xi t}C'_l(0),\;\;\ l=1,\ldots, n-1,$$
$$A'_l(t)\approx A'_l(0),\;\;\;\;\ l=0,1,\ldots, n-1,$$
and so by using (\ref{eq3}), we obtain
$$\hspace{0.5cm}C_l(t)\approx \frac{e^{2i\xi t}}{n}\{e^{-2i\xi nt}\sum_{k=0}^{n-1}C_k(0)+\sum_{k=0}^{n-1}[\sum_{i=1}^{n-1}\omega^{(k-l)i}]C_k(0)\}=
\frac{e^{2i\xi t}}{n}\sum_{k=0}^{n-1}(e^{-2i\xi
nt}-1+n\delta_{k,l})C_k(0),$$
\begin{equation}\label{res}
\hspace{-7.25cm}A_l(t)\approx A_l(0); \;\;\;\;\;\;\;\;\;\
\mbox{for}\;\;\;\;\;\;\ l=0,1,\ldots, n-1
\end{equation}
where, in the first relation we have used the fact that for the
$n$-th root of unity $\omega$, we have
$\sum_{i=0}^{n-1}\omega^{(k-l)i}=n\delta_{k,l}$ and so
$\sum_{i=1}^{n-1}\omega^{(k-l)i}=n\delta_{k,l}-1$. The above
results, are in correspondence with those of Refs.
\cite{Alex1,Alex2} for the special cases $n=2$ and $n=3$. Moreover,
the relations (\ref{res}) indicate that in the limit of large
hopping strength, the state associated with the initially
unentangled atoms, i.e., the initial state with $A_l(0)=0$, for
$l=0,1,\ldots, n-1$, remains effectively unentangled forever.

In the limit of small hopping $\xi\ll$, the equations (\ref{eq1})
and (\ref{eq2}) lead to the following coefficients $C'_i(t)$ and
$A'_i(t)$
$$\hspace{0.25cm}C'_0(t)\simeq e^{-i[1+\xi(n-1)]t}\{[\cos t-i\xi(n-1)\sin t]C'_0(0)-i\sin t A'_0(0)\},$$
$$A'_0(t)\simeq e^{-i[1+\xi(n-1)]t}\{[\cos t+i\xi(n-1)\sin t]A'_0(0)-i\sin t C'_0(0)\},$$
$$\hspace{-1cm}C'_l(t)\simeq e^{-i(1-\xi)t}\{[\cos t+i\xi(n-1)\sin t]C'_l(0)-i\sin t A'_l(0)\},$$
\begin{equation}\label{eq111}
\hspace{2.5cm}A'_l(t)\simeq e^{-i(1-\xi)t}\{[\cos t-i\xi(n-1)\sin
t]A'_l(0)-i\sin t C'_l(0)\};\;\;\;\ l=1,2,\ldots, n-1.
\end{equation}
Now, by neglecting the terms proportional to $\xi$, the above
approximations read as
$$\hspace{0.25cm}C'_0(t)\approx e^{-i[1+\xi(n-1)]t}\{\cos tC'_0(0)-i\sin t A'_0(0)\},$$
$$A'_0(t)\approx e^{-i[1+\xi(n-1)]t}\{\cos tA'_0(0)-i\sin t C'_0(0)\},$$
$$\hspace{-1cm}C'_l(t)\approx e^{-i(1-\xi)t}\{\cos tC'_l(0)-i\sin t A'_l(0)\},$$
\begin{equation}\label{eq111}
\hspace{4cm}A'_l(t)\approx e^{-i(1-\xi)t}\{\cos tA'_l(0)-i\sin t
C'_l(0)\};\;\;\;\ l=1,2,\ldots, n-1.
\end{equation}
Then, by using (\ref{eq3}), one can obtain for $ l=0,1,\ldots, n-1$

$$C_l(t)\approx
\frac{e^{-i(1-\xi) t}}{n}\sum_{k=0}^{n-1}(e^{-i\xi
nt}-1+n\delta_{k,l})(\cos t C_k(0)-i\sin t A_k(0)),$$
\begin{equation}\label{res1'}
A_l(t)\approx \frac{e^{-i(1-\xi) t}}{n}\sum_{k=0}^{n-1}(e^{-i\xi
nt}-1+n\delta_{k,l})(\cos t A_k(0)-i\sin t C_k(0)),
\end{equation}
It could be noted that for times such that $\xi n t\ll 1$, also for
times such that $\xi t=\frac{2k\pi}{n}$ with $k\in Z$, the above
result leads to $C_l(t)\cong e^{-i(1-\xi) t}(\cos t C_l(0)-i\sin t
A_l(0))$ and $A_l(t)\cong e^{-i(1-\xi) t}(\cos t A_l(0)-i\sin t
C_l(0))$, so that we have
$|C_l(t)|^2+|A_l(t)|^2=|C_l(0)|^2+|A_l(0)|^2$ and so, there is no
exchange between the cavities. On the other hand, for the times such
that $\xi t=\frac{(2l+1)\pi}{n}$, with $l\in Z$, the exchange
between the cavities (excitation or photon transfer) can be
achieved. For instance, in the case of two cavities $n=2$, for the
initial state $\ket{\psi(0)}=\ket {a_0}=\ket{e,0}\ket{g,0}$ with
initial conditions $A_0(0)=1$ and $C_0(0)=A_1(0)=C_1(0)=0$, by using
(\ref{res1'}), we obtain at times $t\simeq \frac{(2l+1)\pi}{2\xi}$,
$C_0(t)=A_0(t)=0$, $C_1(t)=-ie^{-i(1-\xi)t} \sin t$ and
$A_1(t)=e^{-i(1-\xi) t}\cos t$, so that we have
$\ket{\psi(t)}=e^{-i(1-\xi) t}(\cos t\ket{g,0}\ket{e,0}-i\sin t
\ket{g,0}\ket{g,2})$.

The results of this section can be used in order to discuss about
qubit state transfer, photon transition and entanglement generation
between the atoms. In order to clarify that, how one can discuss
these arguments, we will consider the special cases of two and three
identical cavities in the next sections in details.
\section{Two coupled cavities: the case $n=2$}
For two cavities ($n=2$), by using the relations
(\ref{eq1})-(\ref{eq3}), one can calculate
$$C_0(t)=\frac{C'_0+C'_1}{\sqrt{2}}=\frac{e^{-it}}{\sqrt{1+\xi^2}}\{[\sqrt{1+\xi^2}\cos \xi t\cos t\sqrt{1+\xi^2}-\xi \sin \xi t\sin t\sqrt{1+\xi^2}]C_0(0)-$$
$$i[\sqrt{1+\xi^2}\sin \xi t\cos t\sqrt{1+\xi^2}+\xi \cos \xi t\sin
t\sqrt{1+\xi^2}]C_1(0)-i\sin t\sqrt{1+\xi^2}(\cos \xi t A_0(0)-i\sin
\xi t A_1(0))\},$$

$$C_1(t)=\frac{C'_0-C'_1}{\sqrt{2}}=\frac{e^{-it}}{\sqrt{1+\xi^2}}\{-i[\sqrt{1+\xi^2}\sin \xi t\cos
t\sqrt{1+\xi^2}+\xi \cos \xi t\sin t\sqrt{1+\xi^2}]C_0(0)+$$
$$[\sqrt{1+\xi^2}\cos \xi t\cos t\sqrt{1+\xi^2}-\xi
\sin \xi t\sin t\sqrt{1+\xi^2}]C_1(0)-i\sin t\sqrt{1+\xi^2}(-i\sin
\xi t A_0(0)+\cos \xi t A_1(0))\},$$

$$A_0(t)=\frac{A'_0+A'_1}{\sqrt{2}}=\frac{e^{-it}}{\sqrt{1+\xi^2}}\{[\sqrt{1+\xi^2}\cos \xi t\cos t\sqrt{1+\xi^2}+\xi \sin \xi t\sin t\sqrt{1+\xi^2}]A_0(0)-$$
$$i[\sqrt{1+\xi^2}\sin \xi t\cos t\sqrt{1+\xi^2}-\xi \cos \xi t\sin
t\sqrt{1+\xi^2}]A_1(0)-i\sin t\sqrt{1+\xi^2}(\cos \xi t C_0(0)-i\sin
\xi t C_1(0))\},$$

$$A_1(t)=\frac{A'_0-A'_1}{\sqrt{2}}=\frac{e^{-it}}{\sqrt{1+\xi^2}}\{-i[\sqrt{1+\xi^2}\sin \xi t\cos
t\sqrt{1+\xi^2}-\xi \cos \xi t\sin t\sqrt{1+\xi^2}]A_0(0)+$$
\begin{equation}\label{eq1n2}
[\sqrt{1+\xi^2}\cos \xi t\cos t\sqrt{1+\xi^2}+\xi \sin \xi t\sin
t\sqrt{1+\xi^2}]A_1(0)-i\sin t\sqrt{1+\xi^2}(-i\sin \xi t
C_0(0)+\cos \xi t C_1(0))\}.\end{equation}

For instance, for the initial state
$\ket{\psi}=\frac{1}{\sqrt{2}}(\ket{c_0}+\ket{c_1})=\frac{1}{\sqrt{2}}(\ket{g,2}\ket{g,0}+\ket{g,0}\ket{g,2})$,
we have the initial conditions $C_0(0)=C_1(0)=\frac{1}{\sqrt{2}}$
and $A_0(0)=A_1(0)=0$. Then, by using the relations (\ref{eq1n2}),
the evolved state of the system will take the form
$$\hspace{-1cm}\ket{\psi(t)}=\frac{e^{-i(1+\xi)t}}{\sqrt{2(1+\xi^2)}}\{[\sqrt{1+\xi^2}\cos \xi t\cos t\sqrt{1+\xi^2}-i\xi \sin t\sqrt{1+\xi^2}](\ket{c_0}+\ket{c_1})-i\sin t\sqrt{1+\xi^2}(\ket{a_0}+\ket{a_1})\}.$$

Now, in order to investigate generation of entanglement between two
atoms, we can evaluate the density matrix $\rho_a$ associated with
the atoms as
$$\rho_a(t)=Tr_c(\ket{\psi(t)}\langle \psi(t)|)=\frac{1}{2(1+\xi^2)}\{2(\xi^2+\cos^2 t\sqrt{1+\xi^2})(\ket{g}^{(1)}\langle g|\otimes \ket{g}^{(2)}\langle g|)+$$
$$\sin^2 t\sqrt{1+\xi^2}(\ket{e}^{(1)}\langle e|\otimes \ket{g}^{(2)}\langle g|+\ket{g}^{(1)}\langle g|\otimes \ket{e}^{(2)}\langle e|+\ket{e}^{(1)}\langle g|\otimes \ket{g}^{(2)}\langle e|+\ket{g}^{(1)}\langle e|\otimes \ket{e}^{(2)}\langle g|)\}$$
where, $Tr_c$ denotes the partial trace over the photonic states
$\ket{2,0}, \ket{0,2}$ and $\ket{0,0}$. Now, for a given hopping
parameter $\xi$, one can use the Peres-Horodecki criteria
\cite{peres1, peres2} known also as positive partial transpose (PPT)
criteria, in order to determine that for which times $t$, the state
$\rho_a(t)$ is entangled, and particularly we can obtain the time
$T$ at which the perfect transfer of photonic entanglement to the
atomic one, is achieved. To this end, we choose the order of atomic
basis as $\ket{g,g}, \ket{g,e}, \ket{e,g}$ and $\ket{e,e}$, so that
the partial transpose of the atomic state takes the following matrix
form
$$(\rho_a(t))^{T_1}=\left(
\begin{array}{cccc}
  2(\xi^2+\cos^2 t\sqrt{1+\xi^2}) & 0 & 0 & \sin^2 t\sqrt{1+\xi^2} \\
  0 & \sin^2 t\sqrt{1+\xi^2} & 0 & 0 \\
  0 & 0 & \sin^2 t\sqrt{1+\xi^2} & 0 \\
  \sin^2 t\sqrt{1+\xi^2} & 0 & 0 & 0 \\
\end{array}\right).
$$
The corresponding eigenvalues of $(\rho_a(t))^{T_1}$ are given by
$\lambda=\sin^2 t\sqrt{1+\xi^2}$ with double degeneracy, and
$\lambda_{\pm}=(\xi^2+\cos^2 t\sqrt{1+\xi^2})\pm \sqrt{(\xi^2+\cos^2
t\sqrt{1+\xi^2})^2+\sin^4 t\sqrt{1+\xi^2}}$. Therefore, the
eigenvalue $\lambda_-$ is clearly negative, except for times
$t=\frac{k\pi}{\sqrt{1+\xi^2}}$, $k\in Z$, where the atomic state
$\rho_a$ is separable. In order to evaluate the amount of
entanglement of the atomic state $\rho_a$, one can calculate the
corresponding concurrence \cite{concurrence}, as
$C(\rho_a(t))=\frac{\sin ^2t\sqrt{1+\xi^2}}{1+\xi^2}$. Then, for
times $T=\frac{(2l+1)\pi}{2\sqrt{1+\xi^2}}$, $l\in Z$, (or in the
suitable units, $T=\frac{(2l+1)\pi}{2\sqrt{1+\xi'^2}}$ with
$\xi'=\frac{2\hbar \xi}{E_0^{+}-E_0^-}$) the maximum value of the
atomic entanglement is achieved and the corresponding concurrence
takes its maximum value $C_{max}=\frac{1}{1+\xi^2}$, where the
atomic density matrix will be maximally entangled for small hopping
$\xi\rightarrow 0$ (see Fig.$1$ for different values of $\xi$).
Moreover, this result indicates that, for large hopping strength
$\xi\rightarrow \infty$, we have $C(\rho_a(t))\rightarrow 0$ and so,
the initially unentangled atoms, remain effectively unentangled for
all the next times.

Now, consider the initial state $\ket{\psi(0)}=\ket{e,0}\ket{g,0}$
with $C_0(0)=C_1(0)=A_1(0)=0$ and $A_0(0)=1$. The equations
(\ref{eq1n2}) give
$$C_0(t)=\frac{-ie^{-it}}{\sqrt{1+\xi^2}}\sin t\sqrt{1+\xi^2} \cos  \xi t,$$
$$C_1(t)=\frac{-e^{-it}}{\sqrt{1+\xi^2}}\sin t\sqrt{1+\xi^2} \sin  \xi t,$$
$$A_0(t)=\frac{e^{-it}}{\sqrt{1+\xi^2}}\{\sqrt{1+\xi^2}\cos t\sqrt{1+\xi^2} \cos  \xi t+\xi \sin t\sqrt{1+\xi^2} \sin  \xi t\},$$
$$A_1(t)=\frac{-ie^{-it}}{\sqrt{1+\xi^2}}\{\sqrt{1+\xi^2}\cos t\sqrt{1+\xi^2} \sin  \xi t- \xi \sin t\sqrt{1+\xi^2} \cos  \xi t\},$$
so that, one obtains
$$|C_0(t)|^2+|C_1(t)|^2=\frac{\sin^2 t\sqrt{1+\xi^2}}{1+\xi^2},$$
$$\hspace{1cm}|A_0(t)|^2+|A_1(t)|^2=\frac{\xi^2+\cos^2 t\sqrt{1+\xi^2}}{1+\xi^2}.$$
Therefore, in the limit of large hopping $\xi$, we have
$|A_0(t)|^2+|A_1(t)|^2\rightarrow 1$ and so the only effective terms
which survive, are $A_0(t)$ and $A_1(t)$.

In addition, one can discuss two photon transfer from the first
cavity to the second one. To do so, we consider the initial state
$\ket{\psi(0)}=\ket{g,2}\ket{g,0}$ with initial conditions
$C_0(0)=1$, $C_1(0)=A_0(0)=A_1(0)=0$. Then, the equations
(\ref{eq1n2}) give
$$C_0(t)=\frac{e^{-it}}{\sqrt{1+\xi^2}}\{\sqrt{1+\xi^2}\cos \xi t\cos t\sqrt{1+\xi^2}-\xi \sin \xi t\sin t\sqrt{1+\xi^2}\},$$
$$C_1(t)=\frac{-ie^{-it}}{\sqrt{1+\xi^2}}\{\sqrt{1+\xi^2}\sin \xi t\cos t\sqrt{1+\xi^2}+\xi \cos \xi t\sin t\sqrt{1+\xi^2}\},$$
$$A_0(t)=\frac{-ie^{-it}}{\sqrt{1+\xi^2}}\cos \xi t\sin t\sqrt{1+\xi^2},$$
$$A_1(t)=\frac{-e^{-it}}{\sqrt{1+\xi^2}}\sin \xi t\sin t\sqrt{1+\xi^2}.$$
Now, for large enough $\xi\gg$, we obtain
$$C_0(t)=e^{-it}\cos 2\xi t,\;\;\ C_1(t)=-ie^{-it}\sin 2\xi t,\;\;\  A_0(t)=\frac{-ie^{-it}\sin 2\xi t}{2\xi}\cong 0,\;\;\ A_1(t)=\frac{-e^{-it}\sin^2\xi t}{\xi}\cong 0. $$
Then, after times $T=\frac{(2k+1)\pi}{4\xi}$ with $k\in Z$, we have
$\ket{\psi(T)}=(-1)^{k+1}ie^{-it} \ket{g,0}\ket{g,2}$ with
$|C_1(T)|^2=1$ and so, two photons of the first cavity are
transmitted to the other cavity, perfectly.
\section{Three coupled cavities: Large and Small hoppings}
The case $n=3$ identical cavities can be considered similar to the
case of $n=2$, by using the relations (\ref{eq1})-(\ref{eq3}). Here
we consider only the limits of large and small hopping $\xi$.
\subsection{Large hopping $\xi\gg$}
In the limit of large $\xi$, we use the equation (\ref{res}) to
obtain
$$C_0(t)\approx \frac{1}{3}\{[C_0(0)+C_1(0)+C_2(0)]e^{-4i\xi
t}+[2C_0(0)-C_1(0)-C_2(0)]e^{2i\xi t}\},$$
$$C_1(t)\approx\frac{1}{3}\{[C_0(0)+C_1(0)+C_2(0)]e^{-4i\xi
t}+[-C_0(0)+2C_1(0)-C_2(0)]e^{2i\xi t}\},$$
$$C_2(t)\approx \frac{1}{3}\{[C_0(0)+C_1(0)+C_2(0)]e^{-4i\xi
t}+[-C_0(0)-C_1(0)+2C_2(0)]e^{2i\xi t}\};$$
$$\hspace{-7cm}A_l(t)\approx A_l(0), \;\;\;\;\ \mbox{for}\;\;\;\;\;\;\;\ l=0,1,2.$$
The above results are in correspondence with those of
Ref.\cite{Alex2}. By considering the initial state
$\ket{\psi(0)}=\ket{c_0}=\ket{g,2}\ket{g,0}\ket{g,0}$ with initial
conditions $C_0(0)=1$, $C_1(0)=C_2(0)=A_0(0)=A_1(0)=A_2(0)=0$, we
obtain the evolved state after time $t$ as
$$\ket{\psi(t)}\approx \frac{1}{3}\{(e^{-4i\xi t}+2e^{2i\xi t})\ket{c_0}+(e^{-4i\xi t}-e^{2i\xi t})(\ket{c_1}+\ket{c_2})\}.$$
Then, the probability of observing two photons at the first cavity
($|C_0(t)|^2$) and that of observing two photons at the two other
cavities ($|C_1(t)|^2=|C_2(t)|^2$), are given respectively by
$$|C_0(t)|^2=\frac{1}{9}[1+4(1+\cos 6\xi t)],\;\;\ |C_1(t)|^2=|C_2(t)|^2=\frac{2}{9}[1-\cos 6\xi t],$$
which indicates that for times $T=\frac{(2k+1)\pi}{6\xi}$, with
$k\in Z$, the corresponding two photons initially located at the
first cavity, are transmitted to one of the other cavities with
equal probability $\frac{4}{9}$.
\subsection{Small hopping $\xi\ll$}
In the limit of small hopping $\xi\ll$, the equation (\ref{res1'})
leads to the following results for three identical cavities:
$$\hspace{-1.5cm}C_0(t)\approx \frac{e^{-i(1-\xi)t}}{3}\{(e^{-3i\xi
t}+2)[\cos t C_0(0)-i\sin t A_0(0)]+(e^{-3i\xi t}-1)[\cos t
(C_1(0)+C_2(0))-i\sin t (A_1(0)+A_2(0))]\},$$
$$\hspace{-1.5cm}C_1(t)\approx \frac{e^{-i(1-\xi)t}}{3}\{(e^{-3i\xi
t}+2)[\cos t C_1(0)-i\sin t A_1(0)]+(e^{-3i\xi t}-1)[\cos t
(C_0(0)+C_2(0))-i\sin t (A_0(0)+A_2(0))]\},$$
$$\hspace{-1.5cm}C_2(t)\approx \frac{e^{-i(1-\xi)t}}{3}\{(e^{-3i\xi
t}+2)[\cos t C_2(0)-i\sin t A_2(0)]+(e^{-3i\xi t}-1)[\cos t
(C_0(0)+C_1(0))-i\sin t (A_0(0)+A_1(0))]\},$$
$$\hspace{-1.5cm}A_0(t)\approx \frac{e^{-i(1-\xi)t}}{3}\{(e^{-3i\xi
t}+2)[\cos t A_0(0)-i\sin t C_0(0)]+(e^{-3i\xi t}-1)[\cos t
(A_1(0)+A_2(0))-i\sin t (C_1(0)+C_2(0))]\},$$
$$\hspace{-1.5cm}A_1(t)\approx \frac{e^{-i(1-\xi)t}}{3}\{(e^{-3i\xi
t}+2)[\cos t A_1(0)-i\sin t C_1(0)]+(e^{-3i\xi t}-1)[\cos t
(A_0(0)+A_2(0))-i\sin t (C_0(0)+C_2(0))]\},$$
$$\hspace{-1.5cm}A_2(t)\approx \frac{e^{-i(1-\xi)t}}{3}\{(e^{-3i\xi
t}+2)[\cos t A_2(0)-i\sin t C_2(0)]+(e^{-3i\xi t}-1)[\cos t
(A_0(0)+A_1(0))-i\sin t (C_0(0)+C_1(0))]\}.$$

Now, by considering for example the initial state
$\ket{\psi(0)}=\ket{a_0}=\ket{e,0}\ket{g,0}\ket{g,0}$, with initial
conditions $A_0(0)=1$ and $C_0(0)=C_1(0)=C_2(0)=A_1(0)=A_2(0)=0$, we
obtain
$$C_0(t)\approx \frac{-i\sin
te^{-i(1-\xi)t}e^{-i(1-\xi)t}}{3}(e^{-3i\xi t}+2),\;\;\;\
C_1(t)=C_2(t)\approx \frac{-i\sin
te^{-i(1-\xi)t}e^{-i(1-\xi)t}}{3}(e^{-3i\xi t}-1),$$
$$A_0(t)\approx \frac{\cos te^{-i(1-\xi)t}}{3}(e^{-3i\xi
t}+2),\;\;\;\ A_1(t)=A_2(t)\approx \frac{\cos t
e^{-i(1-\xi)t}}{3}(e^{-3i\xi t}-1).$$ Therefore, after times
$t=k\pi$, with $k\in Z$, the probability amplitudes $C_0(t)$,
$C_1(t)$ and $C_2(t)$ will be zero and we will have
$A_0(k\pi)\approx\frac{(-1)^{k}e^{-i(1-\xi)k\pi}}{3}(e^{-3i\xi
k\pi}+2)$ and
$A_1(k\pi)=A_2(k\pi)\approx\frac{(-1)^{k}e^{-i(1-\xi)k\pi}}{3}(e^{-3i\xi
k\pi}-1)$. This indicates that, by choosing the hopping strength as
$\xi=\frac{2l+1}{3k}$ with large $k\in Z$ and small $l\in Z$, the
excitation of the first atom located at the first cavity, can be
transmitted with equal probability $\frac{4}{9}$, to one of the
other two atoms.
\section{Conclusion}
In summery, the quantum entanglement properties of $n$ coupled
atom-cavity systems via two-photon exchange interaction, was
analyzed. By employing the photon number operator symmetry of the
Hamiltonian, the corresponding Hilbert space's dimension was reduced
from $3^n$ to $2n$ and then by introducing some suitable Fourier
transformed basis states for the state space of the system, the
corresponding Hamiltonian was block-diagonalized with $2$
dimensional blocks. Due to this useful reduction, the corresponding
Shr\"{o}dinger equation was solved exactly for any number $n$ of
atom-cavities, and excitation and photon transition between the
atoms and the cavities was discussed. The perfect transfer time and
the times at which the maximal quantum entanglement can be
periodically generated between the initially unentangled atoms are
obtained in terms of the hopping parameter (coupling strength)
between the cavities. The large and small hopping limits was
discussed where, it was shown that for large hopping strength, the
initially unentangled atoms remain effectively unentangled forever.

\newpage
{\bf Figure Caption}

{\bf Figure.1:}  Shows the concurrence $C(\rho)$ of the atomic state
$\rho_a(t)$ with respect to time, for different values of hopping
strength (a) $\xi=0.1$, (b) $\xi=0.5$, (c) $\xi=0.9$ and (d)
$\xi=2$.
\end{document}